# Discovery of an extended halo of metal-poor stars in the Andromeda spiral galaxy

Puragra Guhathakurta*, James C. Ostheimer†#, Karoline M. Gilbert*, R. Michael Rich‡, Steven R. Majewski†, Jasonjot S. Kalirai*, David B. Reitzel‡, Michael C. Cooper¶, & Richard J. Patterson†

*University of California Observatories/Lick Observatory, University of California at Santa Cruz, 1156 High Street, Santa Cruz, California 95064, USA

†Department of Astronomy, University of Virginia, PO Box 3818, Charlottesville, Virginia 22903, USA

‡Department of Astronomy, University of California at Los Angeles, Box 951547, Knudsen Hall, Los Angeles, California 90095, USA

¶Department of Astronomy, University of California at Berkeley, Campbell Hall, Berkeley, California 94720, USA

#Present address: Information Extraction and Transport, 1911 North Fort Myer Drive, Suite 600, Arlington, Virginia 22209, USA

**Understanding galaxy formation involves look-back and fossil-record studies of distant and nearby galaxies, respectively[1–4]. Debris trails in our Galaxy's spheroidal halo of old stars provide evidence of "bottom-up" formation via tidal disruption/merging of dwarf satellite galaxies[5–7], but it is difficult to study our Galaxy's large-scale structure from within. Studies of our neighbouring Andromeda galaxy have concluded that its spheroid contains chemically enriched ("metal-rich") stars out to a radius of 30 kiloparsecs with an exponential $r^{1/4}$ fall-off in density thereby resembling a galactic "bulge"[8,9]. Were Andromeda's true halo to be found, our detailed yet global view of its stellar dynamics, substructure,**

**chemical abundance, and age distribution would directly constrain hierarchical halo formation models. We report here on the discovery of a hitherto elusive halo of metal-poor Andromeda stars, distinct from its bulge, with a power-law brightness profile extending beyond $r$ = 160 kiloparsecs. This is 3–5 times larger than any previously mapped Andromeda spheroidal/disk component[10,11]. Together, the Galactic and Andromeda halos span >1/3 of the distance between them, suggesting that stars occupy a substantial volume fraction of our Local Group, and possibly most galaxy groups.**

The Andromeda (M31) inner spheroid was resolved into old "Population II" stars in the mid-1940s[12] but the first investigation of outer spheroid red giant branch (RGB) stars occurred 40 years later[13] and the last two decades have witnessed many key photometric studies[8–11,14,15]. While these studies set out to probe M31's *halo*, the observed properties out to $r = 30$ kpc (for an adopted distance to M31 of 783 kpc) resemble those of the Milky Way (MW) *bulge* instead, albeit a larger/more luminous version (see S.I. for details) — e.g., a "de Vaucouleurs $r^{1/4}$ law" surface brightness profile, $\Sigma(r) = \Sigma_0 \exp[-7.67(r/r_e)^{1/4}]$, and high mean metallicity albeit with a metal-poor tail[9] (the MW halo has an $r^{-2.5}$ brightness profile and ~ 5× lower mean metallicity). M31 has an $r^{-2}$ surface density distribution of globular clusters out to at least $r \sim 20$ kpc[16] and dwarf satellites out to $r \sim 200$ kpc[17], but no associated halo *field* stars were found. M31 appeared to have a single field-star spheroidal component, a bulge, in contrast to the bulge and halo seen in the MW and other spirals.

While M31's proximity makes it easy to *detect* its brightest RGB stars, it is difficult to *identify* them in the galaxy's sparse outer regions against the dense foreground/background of MW dwarf star/galaxy contaminants[18]. During 1998–2001 we imaged nine 36′ × 36′ fields in the *DDO*51 band and Washington system $M$ and $T_2$ bands using the Kitt Peak National Observatory Mayall 4-m telescope and Mosaic camera (Fig. 1)[19]. The intermediate-width $DDO51$ band includes the 517 nm MgH/Mg $b$ spectral absorption feature that is weak (strong) for low (high) surface gravity RGB (dwarf) stars[20]. Each object detected in the Mosaic images was assigned a parameter $f_{DDO51}$ (~ 1/0 for RGB/dwarf stars) based on the fractional overlap of its error ellipse with the known dwarf star locus in the ($M$–$DDO51$) vs. ($M$–$T_2$) colour-colour diagram. We define "M31 RGB candidates" to be objects having $f_{DDO51} > 0.5$ and star-like morphology; the term "candidates" distinguishes them from "confirmed RGB stars" that pass the *L* criterion described below.

Medium-resolution spectroscopy of M31 RGB candidates was carried out with the Keck II 10-m telescope and DEIMOS in 2002–2004[21]. The 1200 lines mm$^{-1}$ grating was used to cover the wavelength range 640–910 nm at a resolution of 0.13 nm (FWHM). Data were obtained in five fields (blue squares in Fig. 1) through 11 multi-slit spectroscopic masks. Each mask has a footprint of ~ 16′ × 4′ and contained 75–100 slits on M31 RGB candidates and lower priority "filler" targets in the brightness range $20 < I_0 < 22.5$. Keck/DEIMOS spectra were also obtained for six DEIMOS multi-slit masks in the inner fields **H11**, **H13s**, and **H13d** (Fig. 1 inset) for which $DDO51$ photometry is not available. Published LRIS data in the inner fields **RG02**[22] and **G1**[23] are also analyzed.

Radial velocities measured from the spectra allow us to identify/remove residual (compact) galaxy contaminants. We have developed a likelihood-based M31 RGB/MW dwarf star separation method using spectral/photometric diagnostics[24]: (**1**) radial velocity, (**2**) $f_{DDO51}$, (**3**) surface-gravity-sensitive 819 nm Na<sub>I</sub> doublet absorption line strength versus $(V–I)_0$ color, (**4**) location in the colour-magnitude diagram (CMD), and (**5**) comparison of the *spectroscopic* metallicity, inferred from the 850 nm Ca<sub>II</sub> triplet absorption line strength, with the *photometric* metallicity, inferred from the CMD location/RGB fiducials. "Training sets" of definite M31 RGB/MW dwarf stars were used to derive empirical probability distribution functions (PDFs) for each diagnostic.

The RGB and dwarf PDFs are distinct from each other, but have some overlap (especially for the CMD diagnostic) implying no single diagnostic is a perfect discriminant (Fig. 2). Each star's measured parameters and corresponding PDFs are used to compute the probability of its being an RGB and dwarf star, $P^i_{giant}$ and $P^i_{dwarf}$, for the $i^{th}$ diagnostic. The overall likelihood is $L = \sum w_i \log(P_{giant}/P_{dwarf})^i / \sum w_i$, where $w_i$ is the weight and the sum runs over the available diagnostics (all five for outer field targets, and all except $f_{DDO51}$ for inner field DEIMOS targets[25]). The stars in each field have a broad/bimodal $L$ distribution with a minimum near 0: those with positive (negative) $L$ values are designated "confirmed" M31 RGB (MW dwarf) stars[24] (see S.I.).

Counts of M31 RGB candidates (point-like objects with $f_{DDO51} > 0.5$) in our full Mosaic data set were converted to surface brightness estimates using the stellar luminosity function of the globular cluster 47 Tucanae[19] (squares in Fig. 3). A constant "background" (dashed line) was subtracted to statistically remove contaminants. The background level was estimated by forcing a match with the confirmed RGB sample at the largest radii (see below); the background uncertainty is ~ 15%[26]. Our two data points at $r$ ~ 20 kpc agree with previous measurements[8] (crosses); these points lie well above the background making them immune to background uncertainties. Beyond $r >$ 50 kpc the surface density of RGB stars relative to contaminants is so low that $DDO$51-based screening alone is not reliable.

M31's surface brightness was also estimated from the confirmed RGB-to-dwarf star count *ratio* times the expected surface density of dwarfs in each field[27] (function of Galactic latitude; see S.I.), with (without) $DDO$51 pre-selection [filled (open) circles in

Fig. 3]. Since we want to measure M31's smooth spheroid, members of the dynamically-cold M31 disk (in fields **H13d** and **G1**) and giant southern stream (in fields **H13s** and **a3**) were statistically subtracted from the total RGB count using fits to the velocity distribution[21,23,25]. The brightness estimates based on *DDO*51 and non-*DDO*51 confirmed RGB stars were normalized separately by forcing agreement with other measurements at overlapping radii (squares/crosses). It has long been known that star counts can probe deeper than integrated surface brightness measurements[8]. Our RGB/dwarf separation method enables us to trace M31's halo to unprecedented brightness levels: $\mu_V \sim 35$ mag arcsec$^{-2}$.

While the previous M31 data were adequately fit by an $r^{1/4}$ law out to $r \sim 30$ kpc[8] (crosses in Fig. 3), our new brightness estimates lie well above the extrapolation of the fit for $r > 50$ kpc (filled circles/squares). This excess is independent of measurement technique and widespread (applies to *all* our outer fields). Our new measurements follow a power-law surface brightness profile, $\Sigma(r) = \Sigma_0 \, r^{-2.3\pm0.3}$, over the range $r = 30$–165 kpc (solid line), similar to the MW stellar halo's profile slope[28]. By analogy with the MW, it seems logical to label the $r^{1/4}$ law and $r^{-2.3}$ power-law components M31's "bulge" and "halo", respectively. Besides having a distinct profile slope, M31's outer halo is significantly more metal poor than its inner spheroid ([Fe/H] $\sim$ –1.5 versus –0.5)[24,29], but comparable to the MW halo[28] (see S.I.).

The three confirmed RGB stars in our outermost **m11** field lie at a projected distance of 50 kpc from M31's companion M33 and their mean velocity is within 100 km s$^{-1}$ of M33's systemic velocity [Fig. 2(**a**)]. Poisson errors and M33's internal motions can account for the latter offset. We cannot exclude the possibility that these three RGB stars are part of a broad M31-M33 tidal bridge[30]. Since M33's scale length and luminosity are smaller than those of M31, it is more likely that these stars are M31 members with $r_{M31} = 165$ kpc rather than M33 members with $r_{M33} = 50$ kpc[13].

The observed spatial and velocity spread ($\sigma_{vel} \sim 100$ km s$^{-1}$) of M31's $r > 50$ kpc RGB stars argues against them belonging to a *single* debris trail. Their velocity and metallicity distributions are consistent with those of M31's present-day dwarf satellites[17]. The large size of M31's halo, $r \sim 165$ kpc, and its observed line-of-sight velocity dispersion implies a *circular* orbital period of nearly 6 Gyr, but the orbits are probably highly eccentric and the period at least 10× smaller.

The halo contains ~ 5% of the total luminosity, assuming it is in a smooth spherical distribution with a kpc-sized constant density core; the fraction may be much higher given the uncertainties in the brightness profile slope/shape/extent, and this could have implications for the light/stellar mass budget of the Universe. The presence of sub-structure (e.g., debris trails[10,11,25,26]) and/or appreciable flattening cannot be ruled out given the limited spatial coverage of our survey. M31's halo structure and dynamics will be better characterized as larger spectroscopic samples become available in the future.

**References (for main manuscript)**

**Acknowledgements**

The $DDO51$, $M$, and $T_2$ imaging data (outer fields) were obtained at Kitt Peak National Observatory of the National Optical Astronomy Observatories, which is operated by the Association of Universities for Research in Astronomy, Inc., under co-operative agreement with the National Science Foundation (NSF). The $g'$ and $i'$ imaging data (inner fields) were obtained with MegaPrime/MegaCam, a joint project of CFHT and CEA/DAPNIA, at the Canada-France-Hawaii Telescope (CFHT) which is operated by the National Research Council of Canada, the Institut National des Science de l'Univers of the Centre National de la Recherche Scientifique of France, and the University of Hawaii. The spectroscopic data were obtained at the W. M. Keck Observatory, which is operated as a scientific partnership among the California Institute of Technology, the University of California and the National Aeronautics and Space Administration (NASA). The Observatory was made possible by the generous financial support of the W. M. Keck Foundation. We are grateful to Marla Geha for help with DEIMOS slitmask designs, to Jim Hesser and Peter Stetson for allowing us to use CFHT/MegaCam data, to Chris Pritchet for providing data in electronic format, and to the DEEP2 Redshift Survey and DEIMOS instrument teams, especially Sandy Faber, Drew Phillips, Alison Coil, Greg Wirth, Kai Noeske, and Jeff Lewis, for help with the instrument, slitmasks, and software. We thank Stephane Courteau for useful discussions about M31's bulge/disk decomposition and Carynn Luine for help with the visual verification of the spectra and absorption line indices. This project was supported by NSF grant AST-0307966 and NASA/STScI grant GO-10265.02 (P.G., K.M.G., and J.S.K.), an NSF Graduate Fellowship (K.M.G.), NSF grants AST-0307842 and AST-0307851, NASA/JPL contract 1228235, the David and Lucile Packard Foundation, and The F. H. Levinson Fund of the Peninsula Community Foundation (S.R.M., J.C.O., and R.J.P.), and NSF grant AST-0307931 (R.M.R. and D.B.R.).

Correspondence and requests for materials should be addressed to P.G. (e-mail: raja@ucolick.org).

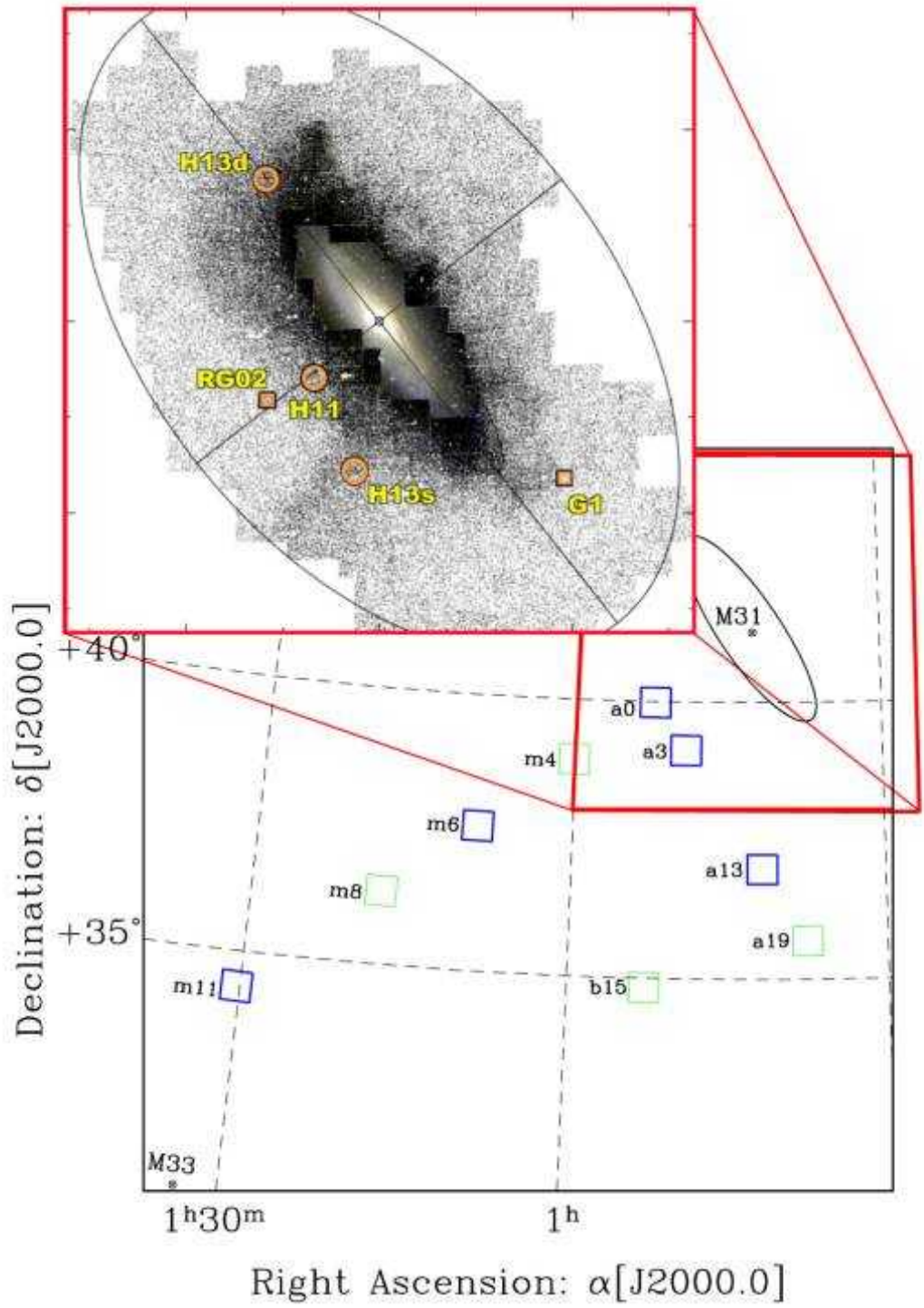

**Figure 1**

**Figure 1** — Location of our fields of study. Blue (green) squares represent Kitt Peak National Observatory 4-m telescope/Mosaic camera images in the intermediate-width ($\Delta\lambda \sim$ 12 nm) *DDO*51 band and Washington *M* and $T_2$ bands[19], with (without) follow-up spectroscopy with the Keck 10-m telescope/DEIMOS[21,24]. Each Mosaic pointing covers 36′ × 36′; the nine pointings span a range of radial distances 2°–12° to the south of M31. The imaging exposure time was ~ 0.25–1.5 hr per filter per field and the seeing FWHM ~ 1″. Standard IRAF data reduction and DAOPHOT point-spread-function fitting techniques were applied to the Mosaic images to obtain photometry for each object (see S.I. for details). The *M* and $T_2$ instrumental magnitudes were transformed to Johnson/Cousins *V* and *I*, respectively, and corrected for reddening by foreground Galactic dust. The 6.5° × 6.5° bold red square around M31 corresponds to the boundary of the inset star-count map[4]. The small orange circles and squares in the inset indicate areas with Keck spectroscopy using DEIMOS[25] and LRIS[22,23], respectively, but with no *DDO*51 photometric pre-selection. For all of our Keck/DEIMOS spectroscopy, with and without *DDO*51 pre-selection, the 1200 lines mm$^{-1}$ grating was used to cover the wavelength range 640–910 nm with a spectral resolution of 0.13 nm. The slit width was 1″, the seeing FWHM 0.8″–1.0″, and the exposure time 1.0–1.5 hr per multi-slit mask. The DEIMOS spectroscopic data were processed using custom-designed software to obtain one-dimensional spectra from which radial velocity estimates were derived via cross-correlation. The CCD mosaic image at the centre of the star-count map shows the traditional view of M31; the ellipse in the larger map represents M31's visible disk and emphasizes how distant our fields of study are from the galaxy's centre. Despite this, we have, to our surprise, found M31 RGB stars in *each* of our fields.

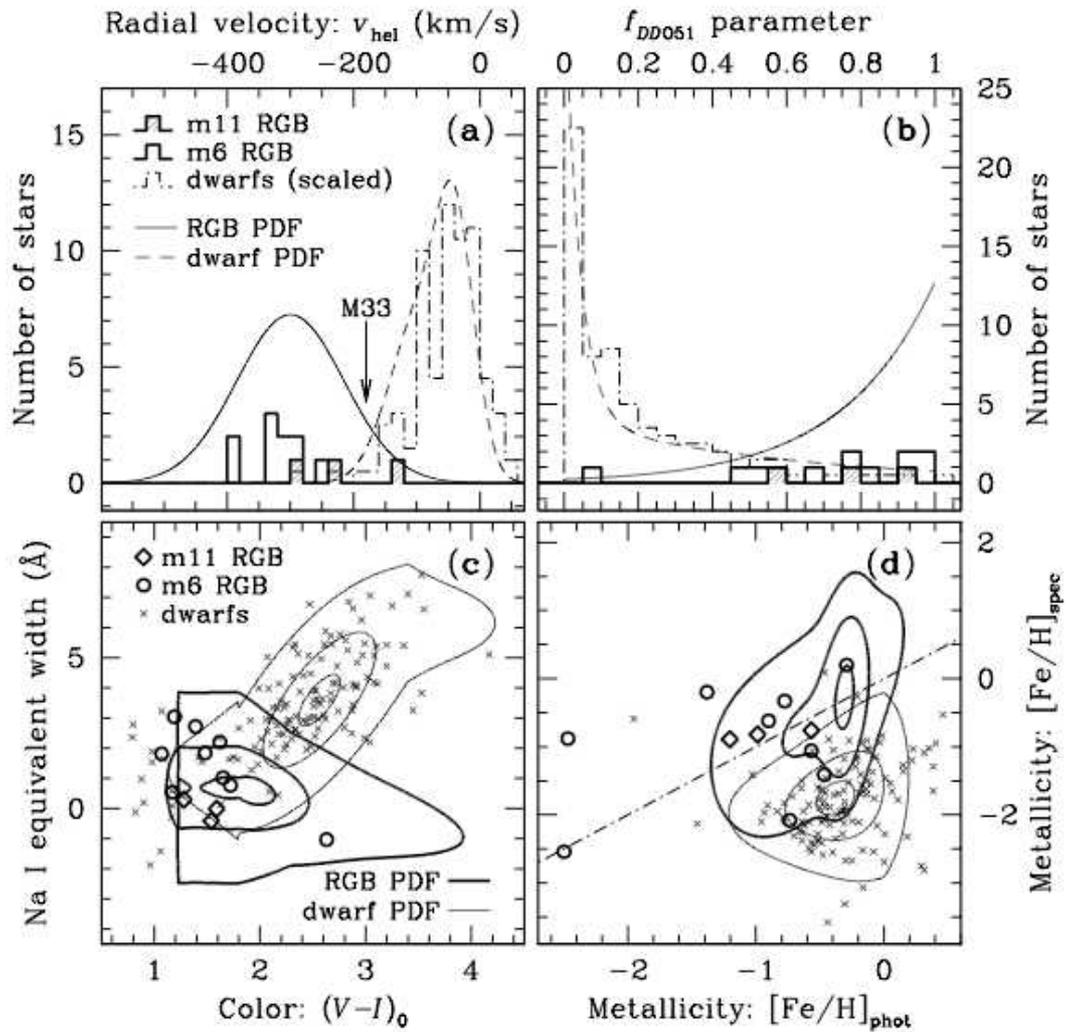

**Figure 2**

**Figure 2** — Illustration of four of the five photometric/spectral diagnostics used to distinguish M31 red giant branch stars from foreground Milky Way dwarf star contaminants[21,24]. (**a**) Radial velocity distribution of confirmed M31 RGB stars in field **m11** (**m6**) [shaded (open) bold histogram] contrasted against that of confirmed MW dwarf stars in these two outermost spectroscopic fields [dot-dashed histogram; scaled down by a factor of 2 to highlight the RGB distribution]. These data are compared to empirical probability distribution functions derived from analytic fits to the distribution of M31 RGB and MW dwarf "training set" stars (solid and dashed curves, respectively). These "training sets" consist of definite M31 RGB and MW dwarf stars drawn from fields close to and far from M31's centre, respectively. The arrow marks the systemic velocity of M31's companion galaxy M33. (**b**) Same as (**a**) for the parameter $f_{DDO51}$, which is based on the star's location in ($M-DDO51$) vs. ($M-T_2$) colour-colour space. (**c**) Equivalent width of the Na I 819 nm absorption band versus de-reddened $(V-I)_0$ colour for field **m11** (**m6**) RGB stars and dwarf stars [bold diamonds (circles) and crosses, respectively]. Bold and thin contours show 10%, 50%, and 90% enclosed fractions for M31 RGB and MW dwarf star PDFs, respectively. (**d**) Same as (**c**) for the photometric metallicity estimate (CMD based) versus spectroscopic metallicity estimate (derived from the equivalent width of the 850 nm Ca II triplet). The dot-dashed diagonal line shows the one-to-one relation. The *x*-axes of the two-dimensional metallicity and colour-magnitude-diagram diagnostics (latter not shown) are closely related but their *y*-axes are distinct: the former is based on the Ca II spectral absorption line strength while the latter is based on broadband photometry. It is reassuring that the confirmed M31 RGB stars follow the RGB PDF and not the dwarf PDF for *all* diagnostics.

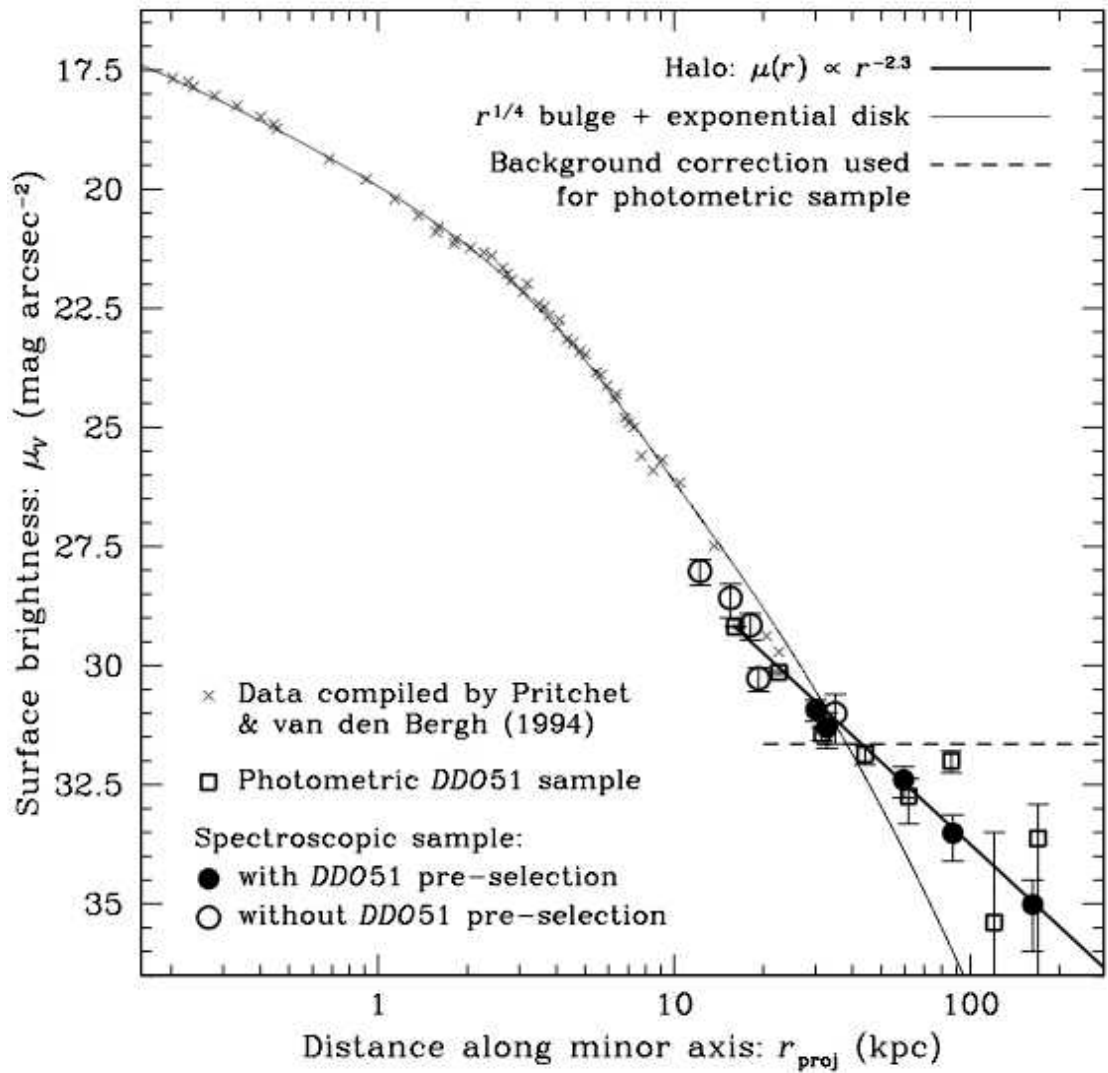

**Figure 3**

**Figure 3** — Surface brightness versus (effective) radial distance along M31's minor axis. Previous integrated surface brightness and star-count-based measurements (crosses) are well fit by the sum of an $r^{1/4}$ law bulge (or, more generally, a Sérsic profile) and an exponential disk (thin solid line) out to $r \sim 20$ kpc and possibly beyond[8]. The squares are based on counts of *DDO*51-selected photometric RGB candidates grouped into radial bins[19]. The dashed line shows the adopted background level, chosen to ensure that the brightness estimates derived from the photometric RGB candidates match those derived from spectroscopically confirmed RGB stars for $r > 50$ kpc[26]. The error bars represent the quadrature sum of Poisson and 15% background subtraction errors. The filled (open) circles are based on confirmed RGB/dwarf star count ratios in our spectroscopic fields with (without) *DDO*51 photometric pre-selection (see S.I.). The spectroscopic field names, in order of increasing radius, are: **H11**, **H13d**, **H13s**, **RG02**, **a0**, **a3**, **G1**, **a13**, **m6**, and **m11**. The RGB count has been statistically corrected for contamination by M31's disk (**H13d** and **G1**) and giant southern stream (**H13s** and **a3**) on the basis of radial velocity data[21,23,25]. This step may add up to a 30% uncertainty over and above the Poisson errors shown. The measurements derived from *DDO*51 and non-*DDO*51 confirmed RGB stars have been normalized independently to match other measurements at small radii. For fields that are not on M31's minor axis (Fig. 1), the effective radial distance along the minor axis was calculated adopting a flattening of $b/a = 0.6$ for $r < 30$ kpc (bulge), as measured for M31's inner spheroid[3,4], and $b/a = 1$ for $r > 30$ kpc (halo). Our brightness estimates are significantly in excess of the extrapolation of the bulge-and-disk fit beyond $r \sim 50$ kpc, but are consistent with an $r^{-2.3}$ power law over the radial range 30–165 kpc (bold solid line).

# SUPPLEMENTARY INFORMATION for the manuscript: Discovery of an extended halo of metal-poor stars in the Andromeda spiral galaxy

**1. Background: Formation of galactic bulges, disks, and halos**

Galaxies come in a rich variety of shapes and sizes. Ellipticals anchor one end of the "Hubble morphological sequence" of galaxies while dwarf irregulars lie at the other end[31,32]. Spirals lie between these two galaxy types in the sequence and contain three subcomponents: disk, bulge, and halo. Bulges and halos are both spheroidal in shape but bulges have a higher central density, a sharper decline in their surface brightness profile with increasing radius in their outskirts ($r^{1/4}$ or Sérsic vs. power-law), and a higher abundance of heavy elements ("metallicity"). Disks form stars at a steady rate over their lifetime; by contrast halo stars [at least in the Milky Way (MW)] formed early in its history and/or were acquired via the accretion of satellite galaxies. The formation and evolution of these galactic subcomponents continues to be the subject of intense theoretical and observational research[33–40].

**2. Previous and ongoing studies of M31's spheroid**

The MW halo contains clues about past merging events[41–46] but our internal vantage point within the disk makes it difficult to map the Galaxy's global properties. Our sister galaxy, the Andromeda spiral (M31), is an ideal target for investigating spheroids because of its relative proximity and highly inclined disk. Early work on the integrated light of M31[47–49] was followed by studies of resolved spheroidal stellar populations over the last two decades[50–61]. These studies suggested that the M31 spheroid out to $r \sim 30$ kpc is a continuation of its inner bulge and resembles the MW bulge (rather than its halo) and elliptical galaxies in terms of the following properties: a de Vaucouleurs $r^{1/4}$ law radial surface brightness profile[53], high metallicity[50,55,59], and high stellar density[57]. Also notable is the presence of substructure in the M31 spheroid and outer disk[60–62].

Spectroscopy of individual red giant branch (RGB) stars in M31[63,64] has shown that the regions sampled in the above photometric studies contain both dynamically hot and cold populations [spheroid (bulge/halo) and disk/stream members, respectively]. Contrary to a recent hypothesis[65], the outskirts of M31 do *not* appear to be totally dominated by a thick disk[66]. In an ongoing study, a coherent rotating structure (disk?) in M31 has been mapped out to an in-plane disk radius of $r_{disk} \sim 40$ kpc[67], but a dynamically hot population is also visible in most, if not all, of the fields in that study. There is also the recent detection of a break/flattening in M31's minor-axis brightness profile at $r \sim 30$ kpc[68].

It may be worth noting here that very low surface brightness stellar halos have recently been detected around a variety of distant edge-on disk galaxies[69,70]. These findings demonstrate that the MW is not alone in possessing a stellar halo. Does its sibling M31 have a stellar halo too?

### 3. Data reduction details

### 3.1 Imaging

Each CCD frame in the KPNO 4-m/Mosaic imaging data set was overscan-subtracted, trimmed, de-biased, and flat-fielded using standard IRAF routines. The resulting frames were then averaged, with cosmic ray rejection, for different exposures of the same field. The object detection, point-spread-function fitting, and photometry software DAOPHOT[71] was run on the combined CCD frames. We classify an object as "star-like" if its DAOPHOT morphological parameters satisfy the criteria: **chi** < 1.3 and |**sharp**| < 0.3. The

United States Naval Observatory astrometric catalogue of bright stars was used to transform the pixel coordinates of the detected objects to RA and DEC.

The KPNO/Mosaic photometry was used to construct ($M$–$DDO$51) vs. ($M$–$T_2$) colour-colour diagrams. High surface gravity dwarf stars tend to form a well-defined locus in this space while low surface gravity RGB stars have a diffuse distribution. The degree of overlap of the photometric error ellipse of each object with the dwarf star locus was used to assign it a "$DDO$51 parameter"; the value of this parameter is ~ 1 for M31 RGB stars and ~ 0 for foreground MW dwarf star contaminants[72–74].

There are five inner fields ($r <$ 20 kpc) in M31 without $DDO$51 data for which we present spectroscopic data (see inset of Fig. 1 of main manuscript). The data for the two LRIS fields have been published[63,75]. The astrometry and photometry on which the DEIMOS spectroscopic multi-slit mask designs for the remaining three fields are based come from two data sets[76]: Canada-France-Hawaii 3.6-m telescope/MegaCam 1 hr imaging exposures covering 1° × 1° in each of the $g'$ and $i'$ bands (fields **H11** and **H13s**) and Keck 10-m/DEIMOS short imaging-mode exposures covering 16′ × 4′ in each of the $V$ and $I$ bands (field **H13d**). These imaging data sets were reduced following the same general procedure described above.

The calibrated magnitudes in the Washington system $M$ and $T_2$ bands and the instrumental magnitudes in the $g'$ and $i'$ bands have been transformed to the Johnson/Cousins $V$ and $I$ bands, respectively, using the following relations:

$$V = M - 0.006 - 0.200\,(M - T_2) \tag{1a}$$
$$I = T_2 \tag{1b}$$
$$V = g' + 0.580 - 0.253\,(g' - i') \tag{1c}$$
$$I = i' - 0.228 - 0.078\,(g' - i') \tag{1d}$$

Equations (1a) and (1b) are from the literature[72] while equations (1c) and (1d) have been derived from CFHT/MegaCam $g'$ and $i'$ observations of Landolt fields containing Johnson/Cousins $V$- and $I$-band photometric standard stars. The apparent $V$ and $I$ magnitudes and colours have been corrected for extinction and reddening to $V_0$ and $I_0$ using the all-sky map of foreground Galactic interstellar dust[77].

**3.2 Spectroscopy**

The Keck/DEIMOS multi-slit spectroscopic data were processed using the SPEC2D software pipelines to obtain sky-subtracted two- and one-dimensional spectra for each targeted object. The SPEC1D pipeline was used to cross-correlate the 1D spectrum of the object against a variety of Sloan Digital Sky Survey stellar, galaxy, and quasar spectral templates in order to obtain a radial velocity estimate[64]. The results were verified using the visual inspection software ZSPEC. The IDL-based SPEC2D and SPEC1D software pipelines and ZSPEC tool were developed by researchers on the DEEP2 galaxy redshift survey team at the University of California at Berkeley (for details see — http://astron.berkeley.edu/~cooper/deep/spec2d/primer.html).

**4. Isolating a clean sample of M31 red giant branch stars**

Our fields of study are located very far from the centre of M31. This is best illustrated in Figure 1 of the main manuscript: the traditional view of the galaxy, represented by a CCD mosaic image[78], shows the bright visible inner disk extending out to a radius of 2.5° ($r \sim 30$ kpc) on the major axis, in contrast to our fields which span the radial range 0.8° – 12° ($r \sim 10 - 165$ kpc) with most, but not all, of the outer fields lying close to M31's south-eastern minor axis. The surface density of M31 RGB stars, relative to that of contaminating foreground MW dwarf stars, is expected to be very low in the outer halo of M31 (only a few percent in our most remote field **m11**). Therefore, it is very important to isolate M31 RGB stars from MW dwarf stars with a high degree of reliability.

**4.1 Diagnostics**

We provide a brief description here of each of the five photometric/spectral diagnostics used in our new likelihood-based RGB/dwarf separation method[64,79] [Fig. 2 of the main manuscript illustrates the details of four of the five diagnostics, all except the colour-magnitude diagram (CMD) based diagnostic (**4**)]:

(**1**) **Radial velocity:** The distribution of training set MW dwarfs is peaked at about $-75$ km s$^{-1}$ with an asymmetric tail towards more negative velocities, similar to what is predicted by the standard Galactic star-count model[80,81]. The MW dwarf velocities tend to be negative because of the reflex of the component of the solar motion in the direction of M31. By contrast, training set M31 RGB stars have a symmetric, broad ($\sigma_{\rm vel} \sim 100$ km s$^{-1}$) distribution of radial velocities centred on the galaxy's systemic velocity of $-300$ km s$^{-1}$.

(**2**) *DDO*51 **parameter:** The M31 RGB and MW dwarf training set stars peak at about 1 and 0, respectively, but each distribution has an extended tail so that the two distributions have substantial overlap.

(**3**) **Equivalent width of Na$_{\rm I}$ doublet at 819 nm:** The Na$_{\rm I}$ doublet is strong in cool (red) dwarf stars but weak in hot (blue) ones and in all RGB stars. In fact, if anything, the formal index value decreases (i.e. becomes more negative) for redder RGB stars.

(**4**) **Colour-magnitude diagram (CMD) location:** The training set M31 RGB distribution follows the shape of the theoretical RGB tracks[82,83] as expected — e.g. the colour spread is small at faint magnitudes but fans out towards the RGB tip, and the RGB tip becomes fainter towards redder colours. By contrast, the MW dwarf distribution does not follow the shape of the RGB fiducials and runs smoothly across the RGB tip. The quantities $X_{CMD}$ and $Y_{CMD}$ used for this diagnostic measure the fractional distance of each object across the tracks (from 0 at the most metal-poor isochrone at [Fe/H] = –2.3 dex to 1 at the most metal-rich one at +0.5 dex) and along them (from 0 at the faint limit of our sample at $I_0 \sim$ 22.5 to 1 at the tip of the RGB), respectively.

(**5**) **Spectroscopic versus photometric metallicity estimate:** The [Fe/H]$_{phot}$ estimate is derived by interpolation between a grid of theoretical RGB tracks[82] in the CMD. The [Fe/H]$_{spec}$ estimate is derived from the weighted sum of the equivalent widths of the Ca$_{II}$ absorption line triplet at 850–866 nm, using calibration relations based on MW globular cluster RGB stars[84]. The latter estimates are relatively noisy, and there could be systematic errors in the above assumptions, but the M31 RGB training set distribution is nevertheless centred roughly on the one-to-one line. MW dwarfs tend to lie well below the line — i.e. they have much weaker Ca$_{II}$ absorption lines.

## 4.2 Probability distribution functions and overall likelihood values

For each of the above five diagnostics, probability distribution functions (PDFs) are constructed by fitting analytic functions to the M31 RGB and MW dwarf training set distributions[79]. Each PDF for the $i^{th}$ diagnostic, $(P_{giant})^i$ or $(P_{dwarf})^i$, is normalized such that the integral under it is unity.

The RGB and dwarf PDFs have a significant degree of overlap with each other for each of the five diagnostics — in other words, none of the above diagnostics by itself is able to discriminate perfectly between these two stellar types. However, the *combination* of diagnostics appears to be an effective RGB/dwarf discriminant. The overall likelihood for each star is defined to be:

$$L = \langle L_i \rangle = \sum w_i \log(P_{\text{giant}}/P_{\text{dwarf}})^i / \sum w_i \tag{2}$$

where $w_i$ is the weight (used to down-weight extreme outliers) and the index $i$ for the summation runs over the available diagnostics. Ignoring the weights for a moment (since they rarely different from unity), the above definition of $L$ can be rewritten as the difference between $L_{\text{giant}}$ and $L_{\text{dwarf}}$, where

$$L_{\text{giant}} = \sum \log(P_{\text{giant}})^i \tag{3a}$$
$$L_{\text{dwarf}} = \sum \log(P_{\text{dwarf}})^i \tag{3b}$$

represent the overall likelihood of the star being an M31 RGB star and MW dwarf, respectively. Thus, a positive $L$ value indicates that the star is more likely to be an M31 RGB star than a MW dwarf star and *vice versa* for a negative $L$ value. Tests show that the diagnostics are more or less independent of one another so that it is appropriate to take the product of the probabilities from the different diagnostics[79].

Figure S1 shows the distribution of overall likelihood $L$ values for each of our Keck/DEIMOS fields arranged in order of increasing projected radial distance from M31. The shaded and open portions of each histogram indicate nominal M31 RGB stars ($L > 0$) and MW dwarf stars ($L < 0$), respectively, and these two populations tend to be concentrated in peaks near $L \sim +1.5$ and $L \sim -1.5$. As expected the strength of the $L > 0$ peak relative to that of the $L < 0$ peak decreases with increasing radial distance. Fields in which the numbers of RGB and dwarf stars are comparable (such as **a0**, **a13**, and **m6**) shown signs of bimodality in the distribution of $L$ values. Note the RGB peak in field **H13s** (**H13d**) is shifted to the right (left) with respect to those in other fields because: [1] the RGB radial velocity distribution in this field is skewed toward negative (positive) values, away from (toward) the MW dwarf distribution, by members of M31's giant southern stream (disk), and [2] the lack of $f_{DDO51}$ information in this field results in the radial velocity diagnostic playing a substantial role in determining the overall $L$ value.

### 4.3 Final classification of stars

We adopt the general criteria $L > +0.5$ and $L < -0.5$ to define confirmed M31 RGB and MW dwarf stars, respectively. The remaining objects are generally considered marginal M31 RGB stars ($0 < L < +0.5$) or marginal MW dwarf stars ($-0.5 < L < 0$). In addition to these $L$ criteria, we consider the CMD location of the star relative to theoretical RGB tracks[52] as measured by its $X_{CMD}$ parameter. If a star is significantly bluer than the most metal-poor of the theoretical RGB isochrones used in our analysis, [Fe/H] = –2.3 (i.e., specifically if $X_{CMD} < -0.05$), it is unlikely to be anything but an MW dwarf star and it is classified as one. A series of statistical tests show that the stars classified as "confirmed M31 RGB stars" and "confirmed MW dwarf stars" do indeed have the properties expected of M31 RGB stars and MW dwarf stars (as judged from the training sets), while both "marginal" categories appear to contain a mix of M31 RGB and MW dwarf stars. As an example, Figure 2 of the main manuscript demonstrates that confirmed M31 RGB stars in fields **m6** and **m11** (bold histograms and symbols) differ from MW dwarf stars (dot-dashed histograms/crosses) in the sense that they follow the RGB PDFs, and not the dwarf PDFs, in terms of *all* of their properties.

### 4.4 Upcoming refinements

Ongoing/future work on RGB/dwarf star separation should lead to some improvement. Other spectral diagnostics are currently being explored including the $K_I$ doublet located at 767 – 770 nm and the TiO bands at 710, 760, and 850 nm. The $K_I$ lines, especially the bluer one, and the 760 nm TiO band are affected by the telluric A-band that is improperly corrected during spectral data processing. The RGB/dwarf star trends in the K I diagnostic plot strongly resemble those in the Na I plot: even though these two diagnostics are not independent, combining them tends to reduce any scatter caused by line strength measurement error. The TiO band is stronger for redder stars and this trend is stronger for RGB stars than dwarf stars. Finally, general-purpose classification methods (such as "neural network" algorithms) may well provide the ideal solution to this RGB/dwarf star separation problem.

## 5. Surface brightness estimates from M31 RGB stars counts

### 5.1 Description of method

A new aspect of our study is the estimation of surface brightness from counts of spectroscopically confirmed M31 RGB stars. This involves the following steps:

(**a**) A *DDO*51-based photometric criterion and a morphological (point-like) criterion is used to pre-select likely M31 RGB star "candidates" for spectroscopic follow up.

(**b**) Only a subset of the M31 RGB candidates within a KPNO/Mosaic field can actually be included in any given Keck/DEIMOS spectroscopic multi-slit mask (conversely, some objects not meeting the *DDO*51 and/or morphological criteria are included to fill up the mask). The usable area of a DEIMOS mask, $16' \times 4'$, is a small fraction of the area of a KPNO/Mosaic field, $36' \times 36'$. Moreover, not all the RGB candidates within the DEIMOS mask area can be accommodated because of "slit conflicts"; the *sampling rate* is less than unity in the crowded inner fields but approaches unity in the sparse outer fields.

(**c**) The resulting Keck/DEIMOS spectroscopy allows definite identification of some "confirmed" M31 RGB stars, MW dwarf stars, and compact background galaxies, but with some degree of *incompleteness*. The failed radial velocity cases tend to be objects with poor spectral signal-to-noise ratio and/or background galaxies for which there are no strong spectral features within our spectral window. The incompleteness fraction varies a bit from one multi-slit mask to another because of variations in spectroscopic observing conditions.

(**d**) The *ratio* of confirmed M31 RGB to MW dwarf stars, multiplied by the surface density of MW dwarfs predicted by the standard Galactic star count model[50,51], is used to estimate M31's stellar surface density. Tests suggest that the sampling rate and degree of incompleteness are similar for M31 RGB and MW dwarf stars in any given mask/field.

(**e**) Since we are interested in measuring M31's spheroid/halo, only the star counts for the dynamically hot component are included in the above ratio.

(**f**) Finally these scaled RGB counts are normalized to a *V*-band surface brightness by forcing agreement with other measurements in the inner fields[53,73].

Since *DDO*51 pre-selection boosts the RGB/dwarf ratio, the normalization of the scaled RGB counts to surface brightness must be done independently for the *DDO*51 and non-*DDO*51 spectroscopic samples. The ratio between these two normalization constants is ~ 5.0, an empirical measure of the efficiency of the *DDO*51 photometric screening procedure. This is in good agreement with earlier independent (and also empirical) estimates of the *DDO*51 screening efficiency[72].

The table below lists the of number (percentage) of different types of objects, grouped into two $I_0$ magnitude bins, in four of our outer fields: *DDO*51-selected M31

| Field name | Mag. range | Total | M31 RGB | MW dwarf | Galaxies |
|---|---|---|---|---|---|
| **a0**  [3 masks] | $20.0 < I < 22.0$ | 114 | 61  (54%) | 8   (7%) | 11 (10%) |
| ($r \sim 30$ kpc) | $22.0 < I < 22.5$ | 59 | 5   (8%) | 0   (0%) | 19 (32%) |
| **a13**  [2 masks] | $20.0 < I < 22.0$ | 67 | 13  (19%) | 6   (9%) | 45 (67%) |
| ($r \sim 60$ kpc) | $22.0 < I < 22.5$ | 30 | 1   (3%) | 1   (3%) | 19 (63%) |
| **m6**  [1 mask] | $20.0 < I < 22.0$ | 15 | 4   (27%) | 4   (27%) | 7 (47%) |
| ($r \sim 85$ kpc) | $22.0 < I < 22.5$ | 19 | 1   (5%) | 0   (0%) | 16 (84%) |
| **m11**  [2 masks] | $20.0 < I < 22.0$ | 29 | 3   (10%) | 4   (14%) | 16 (55%) |
| ($r \sim 165$ kpc) | $22.0 < I < 22.5$ | 30 | 0   (0%) | 2   (7%) | 24 (80%) |

RGB candidates targeted for Keck/DEIMOS spectroscopy, confirmed M31 RGB stars, MW dwarfs, and background galaxies. The number of DEIMOS multi-slit masks observed in each field and the approximate projected distance of each field from M31's centre are also indicated. The "filler" targets are not included in the table.

**5.2 Biases and systematic errors**

Given our empirical normalization method in converting counts of confirmed M31 RGB stars to surface brightness estimates (see previous section), only radially-dependent biases and systematic errors (i.e. those that affect the *slope* of the measured surface brightness profile) can affect the final result. There are two slight biases of opposite sign in our above surface brightness estimates that we have not attempted to correct for:

> (*i*) <u>Filler targets</u> — An outer field mask generally contains a higher fraction of "filler" targets (which fail the DDO51 criterion for RGB stars and/or than an inner field mask and, since many of these filler targets are MW dwarfs, this tends to reduce the M31 RGB/MW dwarf ratio. Thus, the derived radial surface brightness profile is biased in the sense of being steeper than the true profile — i.e. the measured power-law slope is more negative than the true slope. As an illustration, a comparison of surface density estimates based on the M31 RGB/MW dwarf star count ratios in fields **a0** and **m11** yields a power-law slope of –2.23 if filler targets are included, but the slope flattens to –1.88 if filler targets are excluded (as in the table above).

(*ii*) <u>Luminosity function</u> — An outer field mask typically contains a higher fraction of faint targets than an inner field mask. For example, the number of $20 < I_0 < 22$ targets is about twice that of the $22 < I_0 < 22.5$ in the 30–60 kpc fields **a0** and **a13**, whereas these two magnitude ranges contain comparable numbers of targets in the outermost fields **m6** and **m11** (90–165 kpc). The M31 RGB luminosity function rises more steeply towards faint magnitudes than the MW dwarf luminosity function. This tends to bias the measured surface brightness profile in the sense of making it flatter than the true profile. The amount of bias is limited because: (*a*) our spectroscopic targets span a relatively small apparent magnitude range; and (*b*) the radial velocity measurement fails for most stars with $I_0 > 22$, as a result of low signal-to-noise ratio in the continuum (see above table), and this causes our confirmed M31 RGB sample to span an even smaller apparent magnitude range.

Other possible sources of systematic error in our surface brightness estimates include:

(*iii*) <u>Dynamically-cold components</u> — Four of our five non-*DDO*51-selected spectroscopic fields, **H13d**, **H13s**, **RG02**, and **G1** (two of the three DEIMOS fields and both LRIS fields) and one of our five *DDO*51-selected DEIMOS spectroscopic fields, **a3**, contain dynamically-cold populations that were statistically subtracted to infer the number of M31 spheroid RGB stars[63,64,66,75,76] (see Fig. S3). These cold populations are RGB members of M31 disk/giant southern stream or foreground MW dwarfs. In the most extreme cases, errors in the subtraction can cause the M31 spheroid RGB count to be uncertain by as much as 30%.

(*iv*) Spectroscopic data quality — Variations in observing conditions (seeing, transparency, etc) from one mask to another can lead to variations in the completeness fraction for radial velocity measurements. This was compensated to some degree by increasing the total observing time for masks observed under sub-optimal conditions. Nevertheless there are slight field-to-field variations in the faint-end limiting magnitude of the confirmed M31 RGB sample, which in turn leads to slight variations in the fraction of the luminosity function that is included in our statistics [see item (*ii*) above].

(*v*) Photometric data quality — Field-to-field variations in the photometric accuracy and image quality of the *DDO*51 data lead to slight variations in the selection efficiency of M31 RGB candidates and the fraction of contaminating foreground MW dwarf stars and compact background galaxies.

(*vi*) Galactic star-count model — The surface density of MW dwarf stars predicted by the Galactic star-count model[80,81] only enters our final surface brightness estimate in a *relative* sense. Any field-dependent error in the predicted projected density of MW dwarf stars (e.g. due to incorrect Galactic latitude dependence of the model or sub-structure in the MW, say) could affect our results.

Our present sample size is small enough that Poisson errors swamp most, if not all, of the above effects. As described below, data from the Fall 2005 observing season (currently undergoing reduction/analysis) should significantly boost our outer halo sample size. Once a large enough sample of confirmed M31 RGB stars is available, it will be instructive to sub-divide the sample more finely — e.g. in terms of apparent magnitude and *DDO*51 parameter. In future papers, we hope to tackle the considerably more difficult task of measuring the sampling rate and completeness fraction for each field and thereby doing a more direct conversion from counts of confirmed M31 RGB stars to surface brightness.

## 6. Fits to M31's bulge, disk, and halo surface brightness profile

It was noted in a study over a decade ago that the minor-axis surface brightness profile of M31, from the centre out to $r \sim 20$ kpc, is reasonably well approximated by the sum of a de Vaucouleurs $r^{1/4}$ law and an exponential disk[53] (see crosses and model profiles in top panel of Fig. S2). However, the authors also pointed out that, in the radial range $r \sim$ 10–20 kpc, the measured profile is slightly steeper than and falls slightly *below* the $r^{1/4}$ law fit. In contrast to the compilation of heterogenous data at smaller radii[47–49,51,52], the data in this radial range consisted of simple, well-calibrated star counts. This discrepancy between the surface brightness measurements and model profiles is even more noticeable when our halo power-law profile is added to the bulge and disk profiles (bold line).

A recent analysis of the metallicity distribution of M31 RGB stars (and a fit to a large-area *I*-band CCD mosaic image[78]) suggested that the bulge may be better characterized by an $n = 1.6$ Sérsic profile than an $r^{1/4}$ law and that it may be significantly smaller than previously thought[65]. The sum of this (smaller/steeper) bulge profile, exponential disk, and power-law halo are somewhat fainter than the data (middle panel of Fig. S2). Moreover, this recent bulge/disk decomposition predicts that the disk should dominate over the bulge at all locations beyond $r \sim 1$ kpc, a prediction that can be tested with presently available M31 RGB radial velocity data sets[66]. An $n = 2.0$ Sérsic profile and standard effective radius for the bulge plus an exponential disk and power-law halo provide a good fit to the data (bottom panel of Fig. S2).

Irrespective of which of these three sets of fits are adopted, our new surface brightness estimates lie well in excess of the extrapolation of the bulge+disk fit for $r > 50$ kpc (filled circles and squares). In fact, the excess is greater and more significant for Sérsic bulge profiles than for the traditional de Vaucouleurs $r^{1/4}$ law profile.

## 7. Radial velocity and metallicity distributions of M31 spheroid/halo red giants

Figure S3 shows the distribution of radial velocities for confirmed M31 RGB stars in our eight Keck/DEIMOS spectroscopic fields[85]. Each field appears to contain a dynamically hot stellar component. The dashed portion of the histogram indicates, where present, a dynamically cold, non-spheroidal component (e.g., M31 disk or giant southern stream members). The sum of two Gaussians (three, in the case of field **H13s**[76]) has been fit to the data and this is what is used for the statistical subtraction of dynamically cold component(s) in the affected fields before computing the surface brightness from the corrected star counts. Radial velocity cuts are used to exclude the dynamically cold components in order to construct a clean (but admittedly incomplete) sample of M31 spheroid RGB stars for the analysis of the metallicity distribution.

Figure S4 shows the cumulative distribution of metallicities (photometric CMD-based estimates) for confirmed M31 spheroid RGB stars in three broad radial bins: "inner" ($r < 20$ kpc) fields, "intermediate" ($r \sim 30$ kpc) fields, and "outer" ($r > 60$ kpc) fields[79,85]. A clear radial trend is seen in the sense of the mean metallicity decreasing with increasing radius: $\langle [Fe/H] \rangle = -0.5, -0.9,$ and $-1.3$. These mean values are slightly lower than the median values listed in Figure S4, reflecting the asymmetric shape of the metallicity distribution functions (i.e., the presence of a metal-poor tail). Kolmogorov-Smirnov tests indicate that the differences among the metallicity distributions of the three bins are highly significant[85].

For the purposes of comparing the above three radial bins, we have chosen to use the *same* set of theoretical RGB tracks (age $t = 12$ Gyr and $[\alpha/\text{Fe}] = 0$)[82] to derive photometric metallicity estimates for all M31 RGB stars. If M31's stellar halo has an enhanced abundance of light (so-called $\alpha$) elements like the MW halo, $[\alpha/\text{Fe}] = +0.3$, the mean photometric metallicity estimate for the $r > 60$ kpc bin drops to $-1.5$ dex. Spectroscopic metallicity estimates, based on the equivalent width of the Ca$_{\text{II}}$ absorption line triplet, for these confirmed M31 spheroid RGB stars are in good agreement with the photometric estimates; for example, the mean spectroscopic metallicity for the $r > 60$ kpc radial bin is $-1.2$ dex[85].

## 8. Miscellaneous details

### 8.1 M31 versus M33 membership

The discovery of an apparent neutral hydrogen tidal bridge between M31 and M33[86] raises the question: Could our newly detected population of confirmed RGB stars located at large projected distances from M31 be associated with this tidal bridge? The broad spread in radial velocities (Fig. S3) and the widespread spatial extent of the RGB population (these stars are found in all of our fields of study shown in Fig. 1 of the main manuscript) argue against such an association. Even so, it is worth checking whether the RGB stars in our outermost field **m11** might belong to M33, at a projected distance of 50 kpc from this galaxy. There are only a handful of studies of M33's halo[50,87] and these indicate that it is far less luminous/smaller than M31's halo. Thus, these outlying field **m11** confirmed RGB stars are more likely to be members of M31.

### 8.2 Orbital period and enclosed mass

We make a rough estimate of the orbital period of stars in the outermost M31 halo field under the (unlikely, but simplifying) *assumption* of circular orbits. The radial velocity data are consistent with a Gaussian width of about 100 km s$^{-1}$ (Fig. S3). This line-of-sight velocity dispersion translates to a circular speed of about $100(3)^{1/2} = 173$ km s$^{-1}$. At a radius of 165 kpc, this corresponds to an orbital period of 6 Gyr and an enclosed mass of $10^{12}$ $M_\odot$. A careful analysis of the stellar dynamics, with a realistic distribution of orbital anisotropies, should yield more accurate estimates of these quantities[88].

**8.3  Halo extent and luminosity fraction**

During the fall 2005 observing season, Keck/DEIMOS spectroscopy was carried out on *DDO*51-selected M31 RGB candidates in several more fields around M31. This recent supplement includes three new fields in the projected distance range $r \sim 100-150$ kpc (only one of which is on the south-eastern minor axis) and they all appear to contain M31 halo RGB stars[79]. Confirmed M31 RGB stars from the new fall 2005 data set are included in the analysis of the radial metallicity gradient (Fig. S4)[85], but not (yet) in the analysis of the surface brightness profile presented here.

As discussed above, the radius of M31's stellar halo appears to be at least 165 kpc. Stars in our Galaxy's halo have been found beyond a distance of 100 kpc from its centre[45,46]. The sum of the radii of the M31 and MW halos, 260 kpc, is about 33% of the distance between them (783 kpc). Thus, stars in these two large galactic halos occupy at least a few percent of the volume of the Local Group; this volume fraction could be much higher if the full extent of these galaxies is much larger than has currently been mapped. Future studies of the M31 halo will test this hypothesis. If true, this would hint at the presence of a widespread distribution of stars in all galaxy groups.

The M31 halo appears to include about 2%–5% of the overall luminosity of the galaxy. This calculation is based on a best-fit power-law slope of –2.3 and an assumed core radius of 1 kpc. It is worth noting that the detailed form and extent of M31's surface brightness profile remain very uncertain at the present time. A slope of –2.0 corresponds to logarithmic divergence of the total light of the halo; such a slope cannot be ruled out by our data. If a large fraction of M31's total luminosity is indeed distributed in the form of a very low surface brightness extended stellar halo and if this is the norm for all large galaxies, that would in turn have profound implications for the extragalactic background light[89].

**8.4 Age of stars in the halo/bulge**

Recently the Advanced Camera for Surveys on the *Hubble Space Telescope* has been used to carry out ultra-deep imaging and accurate photometry of main-sequence stars in M31[90,91]. At present this is probably the most direct means of measuring ages for stars at these distances. These studies have targeted M31's extended bulge/spheroid (a minor-axis field at $r = 11$ kpc), its giant southern stream, and its disk. About a third of the stars in the bulge and stream fields, the most metal-rich ones, appear to be of intermediate age ($t < 8$ Gyr). It would be instructive to obtain age measurements of M31 halo stars and to compare them to the ages of MW halo stars. Fields beyond $r \sim 30$ kpc on the minor-axis of M31, where the surface brightness of its power-law halo exceeds that of its extended Sérsic bulge, are best suited for studies of the stellar halo.

**References (for Supplementary Information section)**

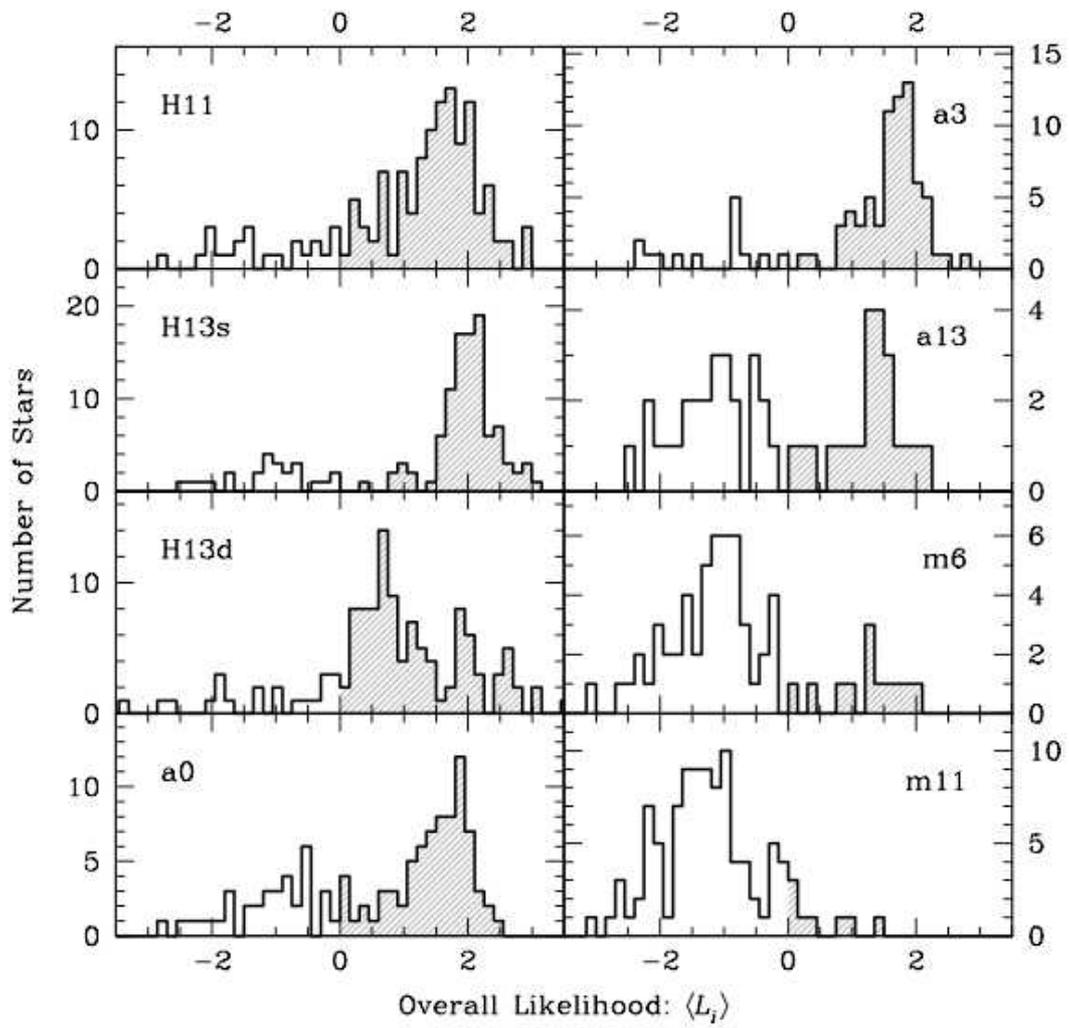

**Figure S1**

**Figure S1** — Distribution of overall likelihood values, $\langle L_i \rangle$, for stars in our eight Keck/DEIMOS spectroscopic fields, arranged in order of increasing projected radial distance from M31's centre[79]. The index $i$ in the weighted average $\langle L_i \rangle$ runs over the five spectroscopic and photometric diagnostics described in Section 4 above. A clear bimodality is seen in most fields: stars with positive $\langle L_i \rangle$ values are generally M31 red giants while the rest are foreground Milky Way dwarfs stars (shaded and open histograms, respectively).

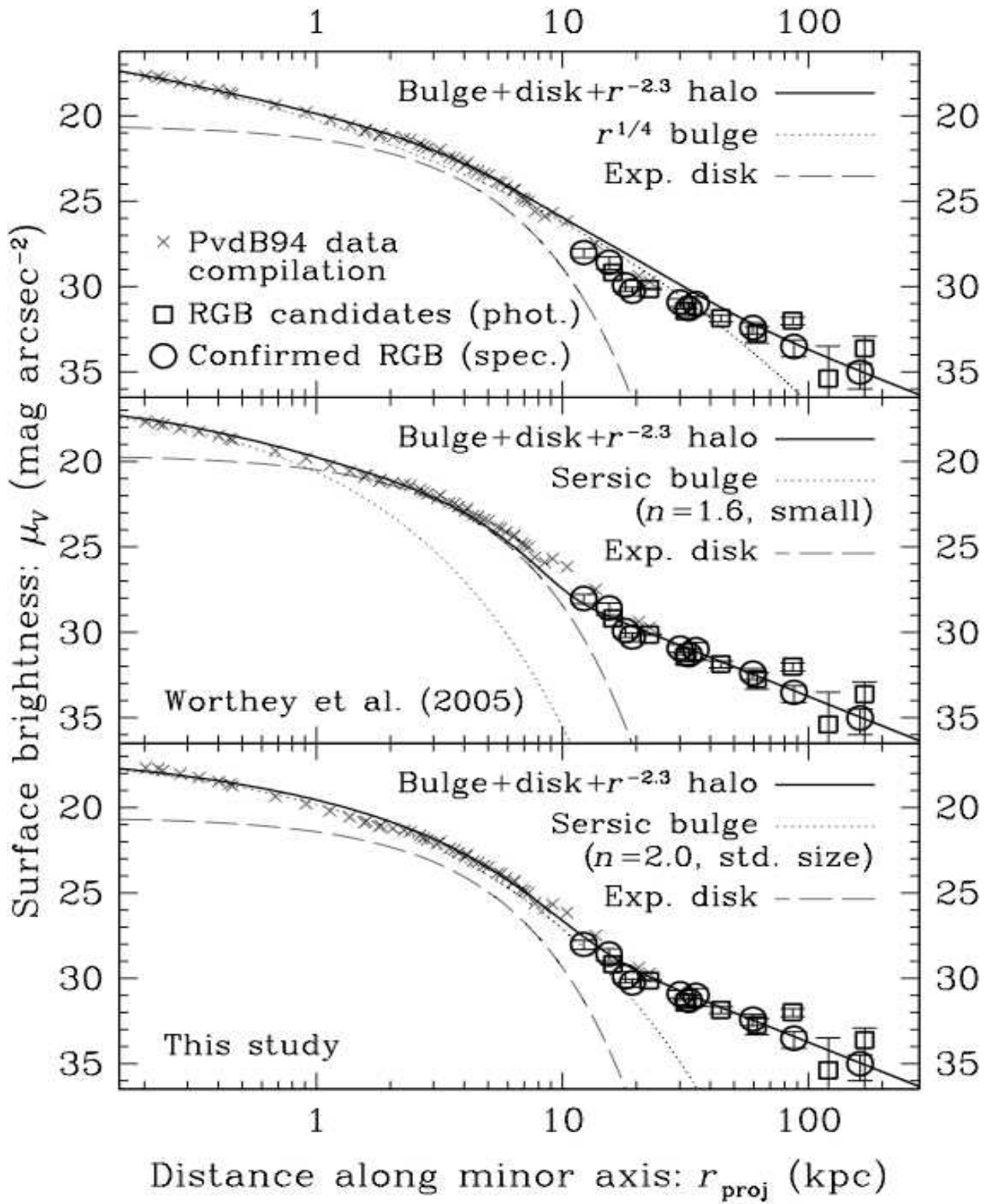

**Figure S2**

**Figure S2** — Surface brightness plotted as a function of (effective) radial distance along M31's minor axis. The data points are the same as those in Figure 3 of the main manuscript (albeit with slightly different symbols; see that figure caption for details): previous integrated brightness and star count-based measurements[53] (crosses) and measurements based on counts of M31 spheroid/halo RGB stars from the photometric and spectroscopic samples in this study (squares and circles, respectively). An $r^{-2.3}$ power-law fit to the halo radial surface brightness profile is combined with three possible decompositions of the bulge (or inner spheroid) and exponential disk (thin dotted and dashed lines, respectively) in the three panels: standard fit using a de Vaucouleurs $r^{1/4}$ law bulge[49] (top); a recently proposed alternate fit using a smaller, $n = 1.6$ Sérsic bulge[65] (middle); and the best-fit from this study using an $n = 2.0$ Sérsic bulge of standard size (bottom). The bold solid line in each panel is the sum of the bulge, disk, and halo model profiles.

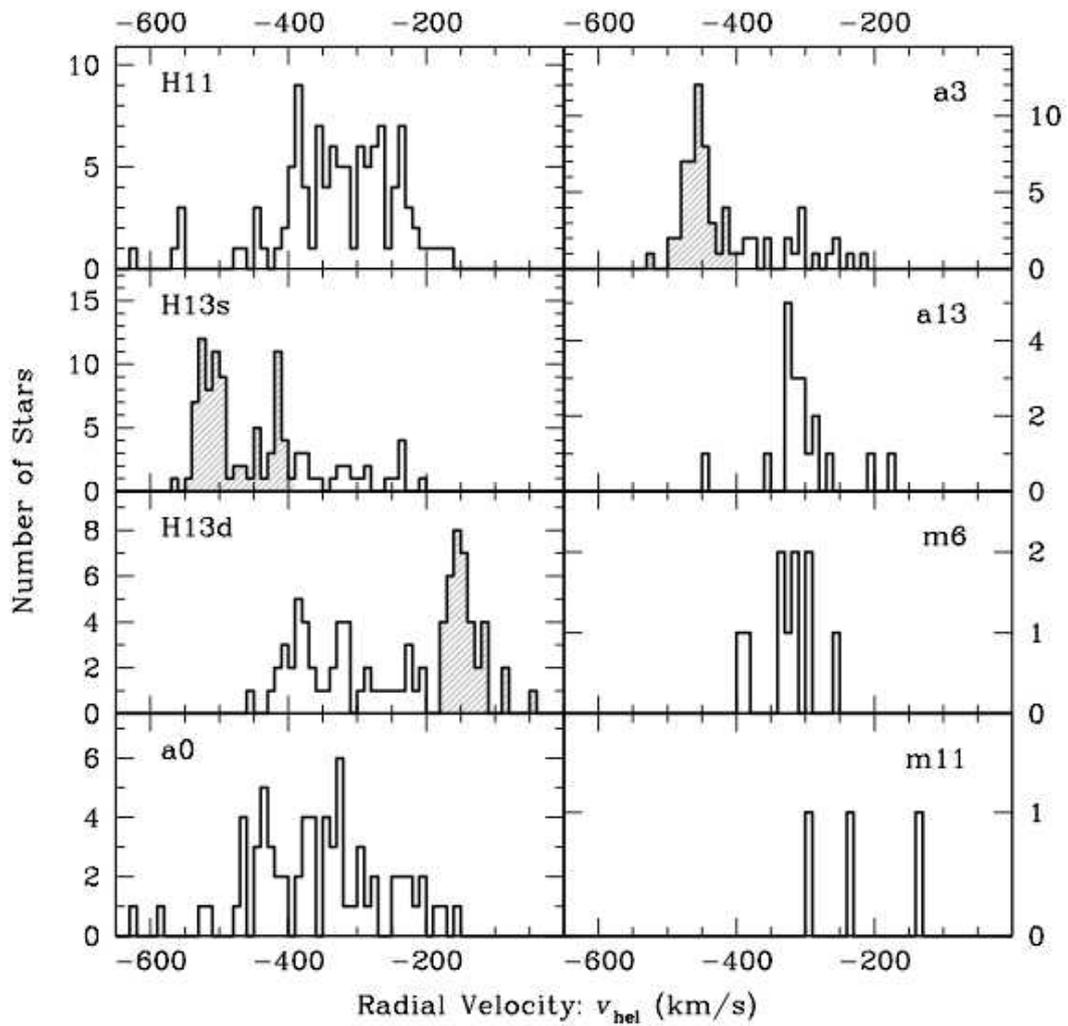

**Figure S3**

**Figure S3** — Radial velocity histograms for confirmed M31 red giant stars in our eight Keck/DEIMOS spectroscopic fields, arranged in order of increasing projected radial distance from M31's centre[85]. For fields that contain two or more dynamical components (**H13d**, **H13s**, and **a3**), multiple Gaussians are fit to the data to separate the dynamically hot halo/spheroid component (open portion of histogram) from the dynamically cold disk and/or satellite debris components (shaded portion of histogram). Fits are not attempted in the three outermost fields due to the paucity of red giants but the data are consistent with the same broad Gaussian that fits the halo/spheroid component in the inner fields.

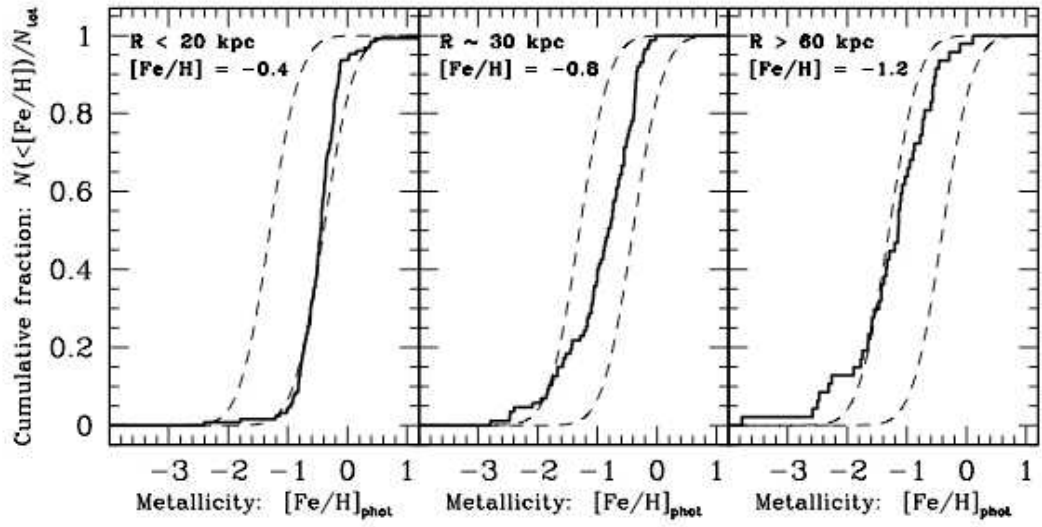

**Figure S4**

**Figure S4** — Cumulative metallicity distribution of M31 halo/spheroid red giants based on photometric estimates[79,85]. The stars have been divided into three groups (left to right): $r <$ 20 kpc (fields **H11**, **H13d**, **H13s**), $r \sim$ 30 kpc (fields **a0** and **a3**), and $r >$ 60 kpc (fields **a13**, **m6**, and **m11**, plus the new Fall 2005 data that include more masks in these fields along with masks in five new outer halo fields). The same pair of Gaussian fiducial curves is shown in all three panels to facilitate comparisons among the three data sets. The median metallicity of each distribution is indicated. There is a statistically significant decrease in the median metallicity with increasing radius.